\documentclass[twocolumn, prb,superscriptaddress]{revtex4-1}
\usepackage{fullpage}
\usepackage{amsmath}
\usepackage{xcolor}
\usepackage{float}
\usepackage{graphicx}
\usepackage[parfill]{parskip}

\begin{document}
\title{A noncollinear density functional theory ansatz for the phononic and thermodynamic properties of $\alpha$-Pu}
\author{Alexander R. Mu\~noz}
\affiliation{T-1: Physics and Chemistry of Materials; Los Alamos National Laboratory, Los Alamos NM 87544}
\author{W. Adam Phelan}
\affiliation{MST-16; Los Alamos National Laboratory, Los Alamos NM 87544}
\author{Matthew S. Cook}
\affiliation{MST-16; Los Alamos National Laboratory, Los Alamos NM 87544}
\author{Greta L. Chappell}
\affiliation{MST-16; Los Alamos National Laboratory, Los Alamos NM 87544}
\author{Paul H. Tobash}
\affiliation{MST-16; Los Alamos National Laboratory, Los Alamos NM 87544}
\author{David C. Arellano}
\affiliation{MST-16; Los Alamos National Laboratory, Los Alamos NM 87544}
\author{Derek V. Prada}
\affiliation{MST-16; Los Alamos National Laboratory, Los Alamos NM 87544}
\author{Travis E. Jones}
\affiliation{T-1: Physics and Chemistry of Materials; Los Alamos National Laboratory, Los Alamos NM 87544}
\author{Sven P. Rudin}
\affiliation{T-1: Physics and Chemistry of Materials; Los Alamos National Laboratory, Los Alamos NM 87544}

\begin{abstract}
Plutonium's phase diagram is host to complex structures and interactions that make the description of its ground state properties elusive.
Using all-electron density functional theory, we study the thermodynamic properties of $\alpha$-Pu.
To do this, we build on recent work in the literature by introducing a novel noncollinear magnetic ansatz for $\alpha$-Pu's ground state.
The noncollinear ansatz accurately recovers the experimental phonon density of states, heat capacity, and thermal expansion.
These new results on $\alpha$-Pu along with recent results on $\delta$-Pu demonstrate the efficacy of noncollinear ansatzes for the description of plutonium.
\end{abstract}

\maketitle
\section{Introduction}
The combination of spin-orbit coupling and electronic correlations makes it challenging to describe the 5f metals.\cite{wallace_1998,clark_2019}
Of the actinides, plutonium presents a particularly tricky case because it has both itinerant and localized electronic behavior.\cite{wills_2000,eriksson_1999,soderlind_2003,boring_2000}
The case of $\alpha$-Pu is of particular interest due to its complex crystal structure that leads to site-selective electronic correlations.\cite{huang_2020,zhu_2013,espinosa_2001}
The variation in the local environments of each Pu atom makes understanding the structural properties of $\alpha$-Pu highly dependent on the electronic structure approach used to describe the system.\cite{xu_2008,svane_2007,penicaud_1997,soderlind_2019_2}

Density functional theory (DFT) has shown mixed results when used  to compute the thermodynamic and phononic properties of many Pu phases.\cite{soderlind_1997,soderlind_2004,soderlind_2009,soderlind_2010}
Agreement with experiment is strongly tied to assumptions of the electronic structure.
For instance, a collinear antiferromagnetic state has been used to approximate the experimentally observed paramagnetic state in $\delta$-Pu.\cite{soderlind_2019,janoschek_2015}
However, in the case of $\delta$-Pu, the collinear state introduces a mechanical instability that can be amended by the introduction of a noncollinear ansatz for the ground state.\cite{rudin_2022,soderlind_2023}
This state, the 3Q magnetic state, fixes the instability by rendering all of the bond directions symmetry equivalent in their magnetic orientation and character.

The use of noncollinear states in DFT is a useful heuristic for approximating the paramagnetic ground state of Pu, as it lowers the energetic gap between neighboring states.\cite{lashley_2005,sakuma_2000,rudin_2022,soderlind_2023}
This approach circumvents the use of cost-prohibitive methods while maitaining the appropriate interactions needed to capture the structural properties of Pu.\cite{pourovskii_2007,zhu_2013,amadon_2016}
For face-centered cubic PuO$_2$, the 1Q, 2Q, and 3Q states all occupy the low-energy noncollinear space, but the state space for more complex crystal structures is substantially larger.\cite{pegg_2018} 
There are techniques for exploring the noncollinear state space, but they are prohibitively expensive for applications to Pu.\cite{payne_2018}  
The identification of noncollinear descriptions is tricky for systems with low symmetry, where a bond-equivalent picture does not intuitively emerge.

In this manuscript, we introduce a noncollinear ansatz for the ground state of $\alpha$-Pu that is derived from the collinear state in the literature.\cite{soderlind_2019}
Using DFT and the small-displacement method for forces, we compute the phonon density of states (DOS) to make comparisons to experimental results including the thermal expansion and specific heat.\cite{manley_2009,schonfeld_1996}
We report heat capacity measurements on a freshly prepared plutonium sample to control for aging and validate the results of the thermodynamic analysis.
The atom-projected electronic and phononic DOS are used to understand the localization of charge in noncollinear $\alpha$-Pu that leads to improvements in the thermodynamic description over the collinear state.
In short, we show that the noncollinear approach to Pu allotropes can be extended into systems where the crystal symmetry obfuscates a bond equivalent picture, namely $\alpha$-Pu.

\begin{figure*}
\centering
\includegraphics[width=6in]{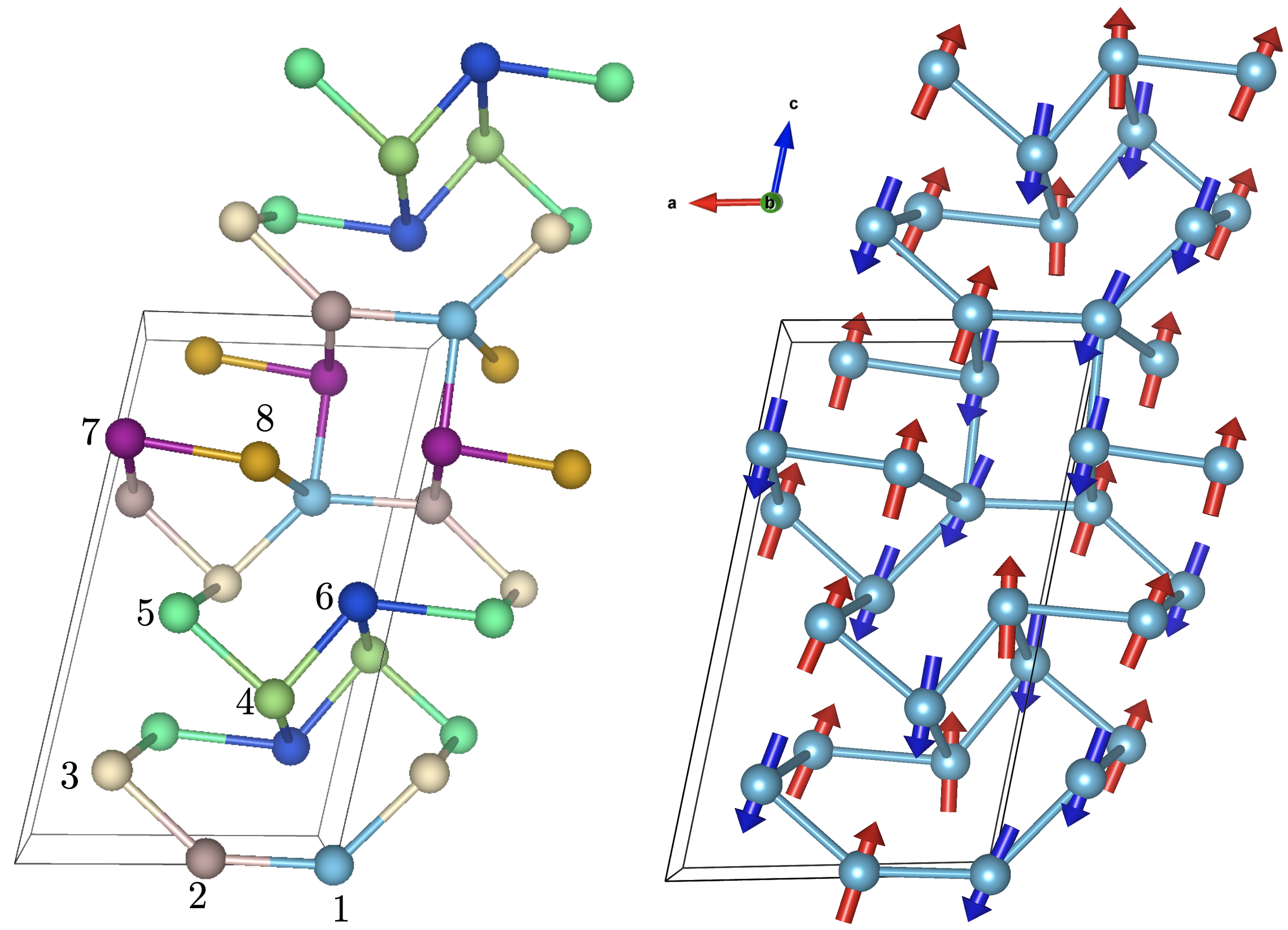}
\caption{Left: Visualization of the eight crystallographically distinct Pu sites in $\alpha$-Pu's unit cell. Bonds are drawn between atoms with distances less than 2.8 $\text{\AA}$, indicating short bonds in the system. Right: The optimized magnetic structure for $\alpha$-Pu. The teal spheres are the Pu atoms with short bonds drawn. The arrows show the direction of the magnetic moments on each site. The color of the arrows indicates a moment in the $+c$ (red) or $-c$ (blue) directions. Visualizations done in VESTA.\cite{vesta}}
\label{fig:structure}
\end{figure*}

\section{Methods}
\subsection{Theory}
The calculations presented in this work were performed using the full-potential linearized augmented plane wave package Elk.\cite{elk}
The functional used in this study is the generalized gradient approximation of Perdew et al. (PBE).\cite{perdew_1996}
We used a $6\times6\times6$ \textbf{k}-point grid with a $\frac{1}{8}\times \frac{1}{8} \times \frac{1}{8}$ offset for the 16-atom $\alpha$-Pu unit cell with the experimental lattice parameters.\cite{zachariasen_1963}
The energy convergence criterion was set to $1\times10^{-4}$ eV and the force convergence was set to $5\times10^{-7} \ \frac{\text{eV}}{\text{\AA}}$.  

The unit cell of $\alpha$-Pu is exceptionally complicated.
There are eight symmetry unique atomic sites in the 16 atom monoclinic cell of $\alpha$-Pu, Figure~\ref{fig:structure}.
The atoms are located in two planes in the b-direction, but the local structures are very complicated.
The bonds drawn in Figure~\ref{fig:structure} show the bond lengths that are less than 2.8 $\text{\AA}$, referred to as short bonds.\cite{zhu_2013}
Atoms 2 through 7 have four neighboring short bonds, two in the a-c plane and two in the b-direction.
Atom 1 has five neighboring short bonds giving it the most constrained environment in the cell.
On the other hand, Atom 8 has three short bonds and sits in the most open environment in the cell. 

The phonon results were obtained using PHON and the small-displacement method.\cite{alfe_2009,alfe_2001,kresse_1995}
After ionic relaxation, the symmetry inequivalent atoms in the computational cell were displaced in all symmetry-inequivalent directions to construct the force constants.
Using PHON, we construct the dynamical matrix for the wave vectors, \textbf{q}.
The dynamical matrix is diagonalized to obtain the phonon frequency at each \textbf{q}.
We sample the Brillouin zone using an evenly spaced $21\times21\times21$ grid, to ensure the convergence of the results within PHON.
We confirmed the convergence of the phonon results by varying the \textbf{k}-point grid ($4\times4\times4$ to $6\times6\times6$) and by using both 16-atom and 32-atom cells for the calculations.

The noncollinear magnetic state was identified by perturbing the AFM magnetic state used in the literature.\cite{soderlind_2019}
Before starting a self-consistent field calculation, each magnetic moment was randomly rotated and given no constraint on its magnitude or direction. 
This procedure was done 20 times to obtain a set of noncollinear states.
The lowest energy magnetic structure is shown in Figure~\ref{fig:structure}.
Magnetic moment constraints were not applied for small-displacements, to allow bonds to soften.

We compute the thermal expansion of $\alpha$-Pu using the quasiharmonic approximation.
Using the computed phonons, we evaluated the free energy at a series of five volumes at temperatures from 0 K to 300 K. 
The free energies were fit to obtain isotherms with an equilibrium volume at a given temperature.
Collating the equilibrium volume at each temperature, we measure the expansion of the volume as a function of temperature.

\subsection{Experiment}
\subsubsection{Materials Synthesis}
A coupon of $\alpha$-Pu was arc-melted in a zirconium-gettered argon atmosphere in order to allow the helium from radioactive aging to be released.
Simultaneously, this resets the defects formed from the radioactive decay.
Prior to arc-melting, the reaction chamber was pumped using a vacuum and purged with argon gas several times to ensure minimal oxygen and moisture content. 
After arc-melting, the resulting button was wrapped in zirconium foil and sealed into an evacuated fused silica tube.
This reaction ampoule was then placed into a box furnace and held at 460 degrees C for one month.
The reaction vessel was then furnace cooled.

\subsubsection{Heat Capacity Measurements}
Heat capacity data was collected on an annealed $\alpha$-Pu piece cut from the arc-melted/annealed button.
The heat capacity measurements were done using a 14-Tesla Quantum Design Dynacool System with a semi-adiabatic pulse technique.
The contribution of n-grease that was used to affix the sample to the sapphire stage was subtracted via a common addenda measurement on a Quantum Design heat capacity puck.

\begin{figure}
\centering
\includegraphics[width=3in]{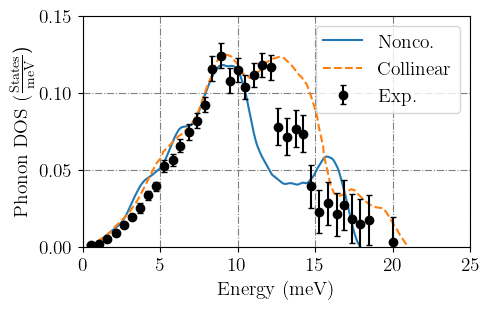}
\caption{The phonon density of states from x-ray scattering compared to the collinear phonon DOS in the literature and the noncollinear state from this manuscript. \cite{manley_2009,soderlind_2019} Notably, the noncollinear result captures a plateau feature in the phonon DOS that is present in the experiment. There is also a reduction in the difference with respect to experiment compared to the collinear phonon DOS. The 12-14 meV region shows a discrepancy in both magnetic structures. }
\label{fig:dos}
\end{figure}

\section{Results}
\subsection{Magnetism and Phononics}
The low-energy magnetic structure, shown in Figure~\ref{fig:structure}, has a net spin magnetic moment near 0 like the experimental description of $\alpha$-Pu.\cite{lashley_2005} 
The remainder of the spin magnetic moment is cancelled out by the orbital magnetic moment on each site, as observed in the DFT literature.\cite{soderlind_2019_2}
On an atom by atom basis, the spin magnetic moments vary from $2\ \mu_B$ to $3\ \mu_B$
The magnetic structure is mainly composed of approximately antiferromagnetic bonds, with significant canting in the $a$-direction.
The ferromagnetic bonds have a significant amount of canting, avoiding the high energetic costs from localization due to Fermi holes.
As such, the noncollinear antiferromagnetic order allows for the redistribution of charge around the crystallographically distinct sites in the lattice.
\begin{figure}
\centering
\includegraphics[width=3in]{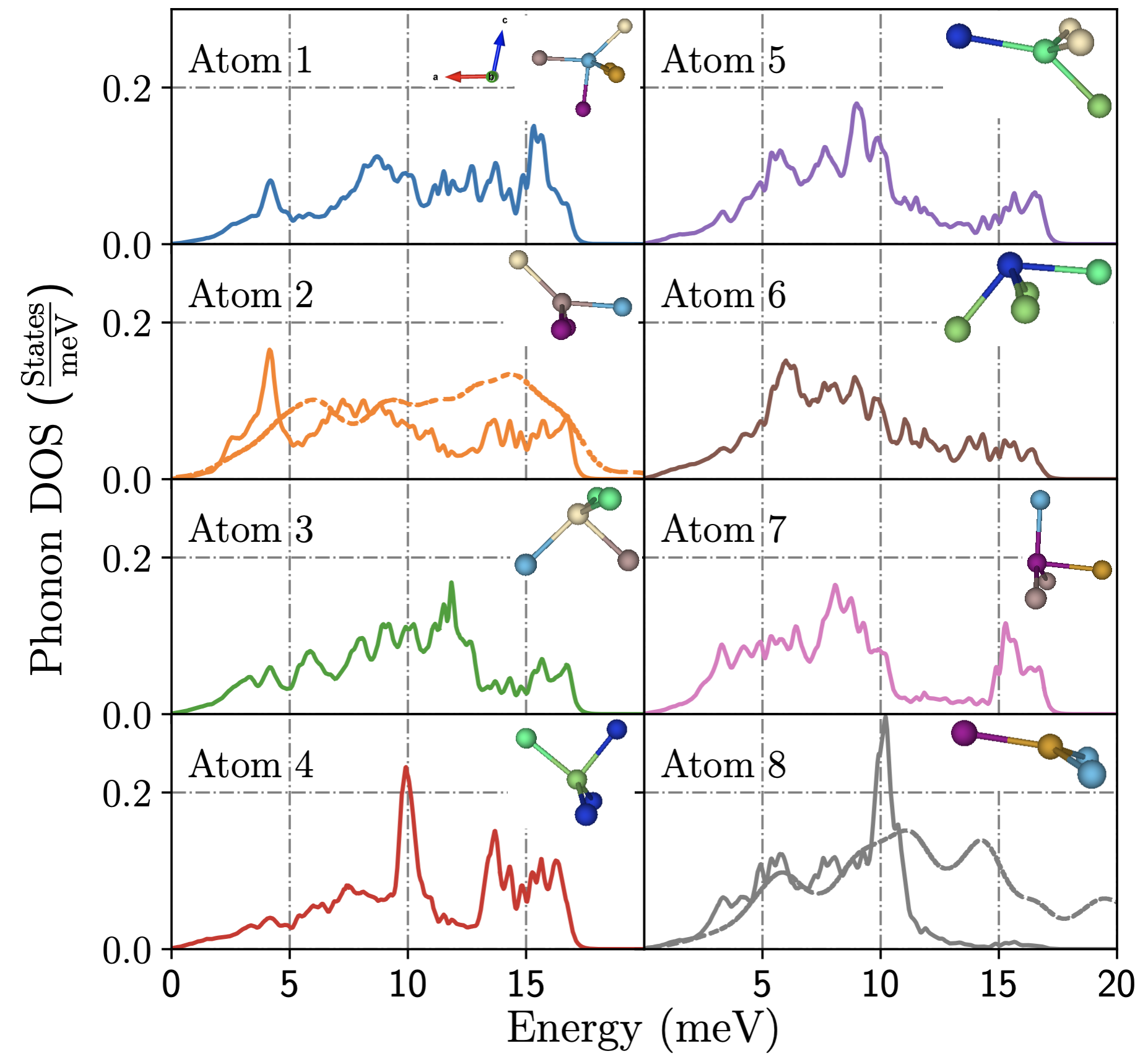}
\caption{The local atom-projected phonon density of states for each atomic type in $\alpha$-Pu. The local environment dictates the shape of the atom-projected phonon DOS, particularly, we observe distinct line shapes on sites 1 and 8. When compared to the literature, dashed lines on Atom 2 and 8, we see a notable decrease in the DOS at energies greater than 10 meV.\cite{soderlind_2019}}
\label{fig:phonon-pdos}
\end{figure}

The phonon density of states is shown in Figure~\ref{fig:dos}.
We give a comparison between the phonon DOS of the collinear state from the literature, our noncollinear state, and phonon DOS from inelastic x-ray scattering.\cite{soderlind_2019,manley_2009}
The noncollinear state alters the description of the phonon DOS with respect to the collinear state, but improves agreement with experiment.
The 12-13 meV window presents the largest discrepancy with experiment in both theoretical approaches.
The experimental DOS features a step in this energetic window which is present in the DOS of the noncollinear state, giving a reasonable agreement at the qualitative level.
The relative softness of the noncollinear state's spectrum indicates the flexibility of the charge redistribution enabled by the noncollinear state. 
However, we expect each atom to behave distinctly due to the differences in their local enviroments. 

In Figure~\ref{fig:phonon-pdos}, we show the local atom-projected phonon DOS for each crystallographically distinct site in the unit cell.
With eight types of atoms there are eigh distinct local environments that influence the phonon DOS at each site.
We show the local environments as subplots in Figure~\ref{fig:phonon-pdos}.
Despite the variations in local environments, there are key patterns for identifying similar behaviors.
Atoms 1 and 2 have peaks at low energy, while Atoms 4 and 8 have strong peaks at 10 meV.
Atoms 4, 5, and 7 have low DOSs in the 10-12 meV window, but increase at higher energies.
Additionally, Atoms 2 and 8 decrease the DOS at high energies with respect to the collinear state.
In combination, these behaviors indicate how we obtained the step-like feature from 10-12 meV that is observed in experiment. 

To study the local structure's influence on the phonon DOS, we examine the phonon DOS for Atoms 1 and 8.
In the case of Atom 8, which has three neighbors at short distances, the DOS abruptly drops at 11 meV while Atom 1, which has five neighbors at short distances, has states spread across the entire range of energetic values.
While Atom 8 is still distinct from other atoms, our result differ significantly from previous DFT calculations which show Atom 8 having a high DOS at high energies.\cite{soderlind_2019}

The presence of neighbors at short distances often stiffens the phonon spectrum, but, in our case, all of Atom 8's neighbors are antiferromagnetically aligned. 
With noncollinear spins, the magnetization can react to shortening bonds for less energetic cost, effectively softening the bonds.
The local environments in Figure~\ref{fig:phonon-pdos} show that an increased DOS at high energies is correlated with bonds in the c-direction.
This points to the presence of stiff bonds in the c-direction.
In contrast, Atom 8 has bonds in the a-b plane and a reduction of stiff phonon modes.
This is partly due to the noncollinear magnetization, and the relatively long lengths of bonds in the b-direction.

\begin{figure}
\centering
\includegraphics[width=3in]{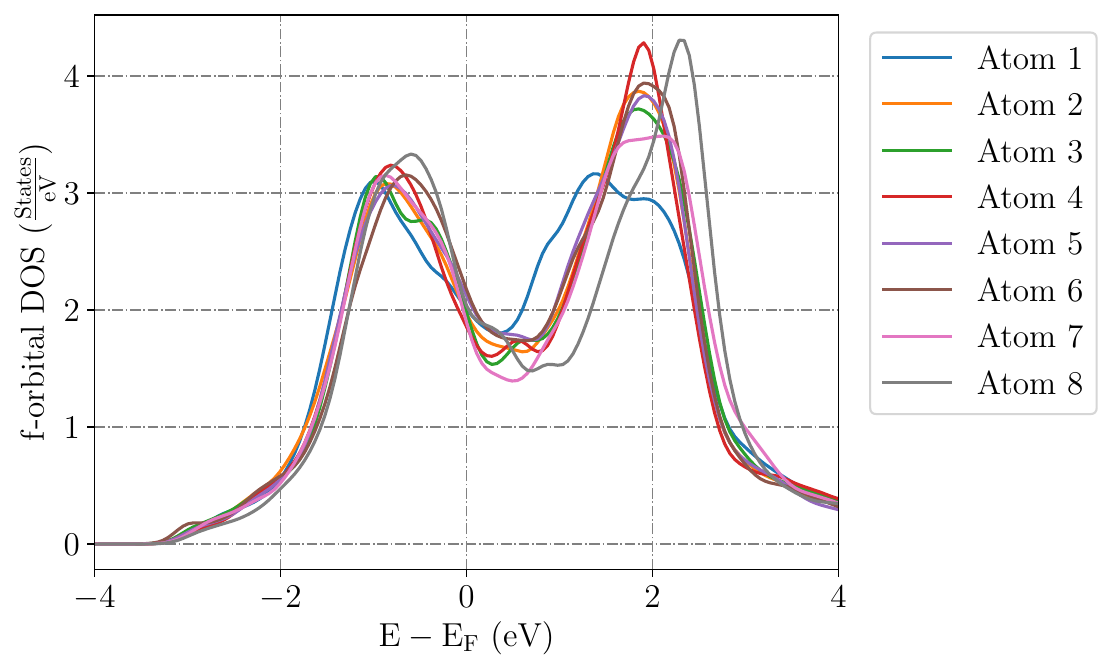}
\caption{The electronic density of states for the eight crystallographically nonequivalent atoms in $\alpha$-Pu. The site dependence of the DOS demonstrates the impact of the local environment on the electronic properties of the material. Atom 1 and Atom 8 have distinct DOS characters, but the local environments around each of the other Pu atoms lead to changes in the charge structure that make them distinct in the electronic DOS.}
\label{fig:e-dos}
\end{figure}

\begin{figure*}
\centering
\includegraphics[width=6in]{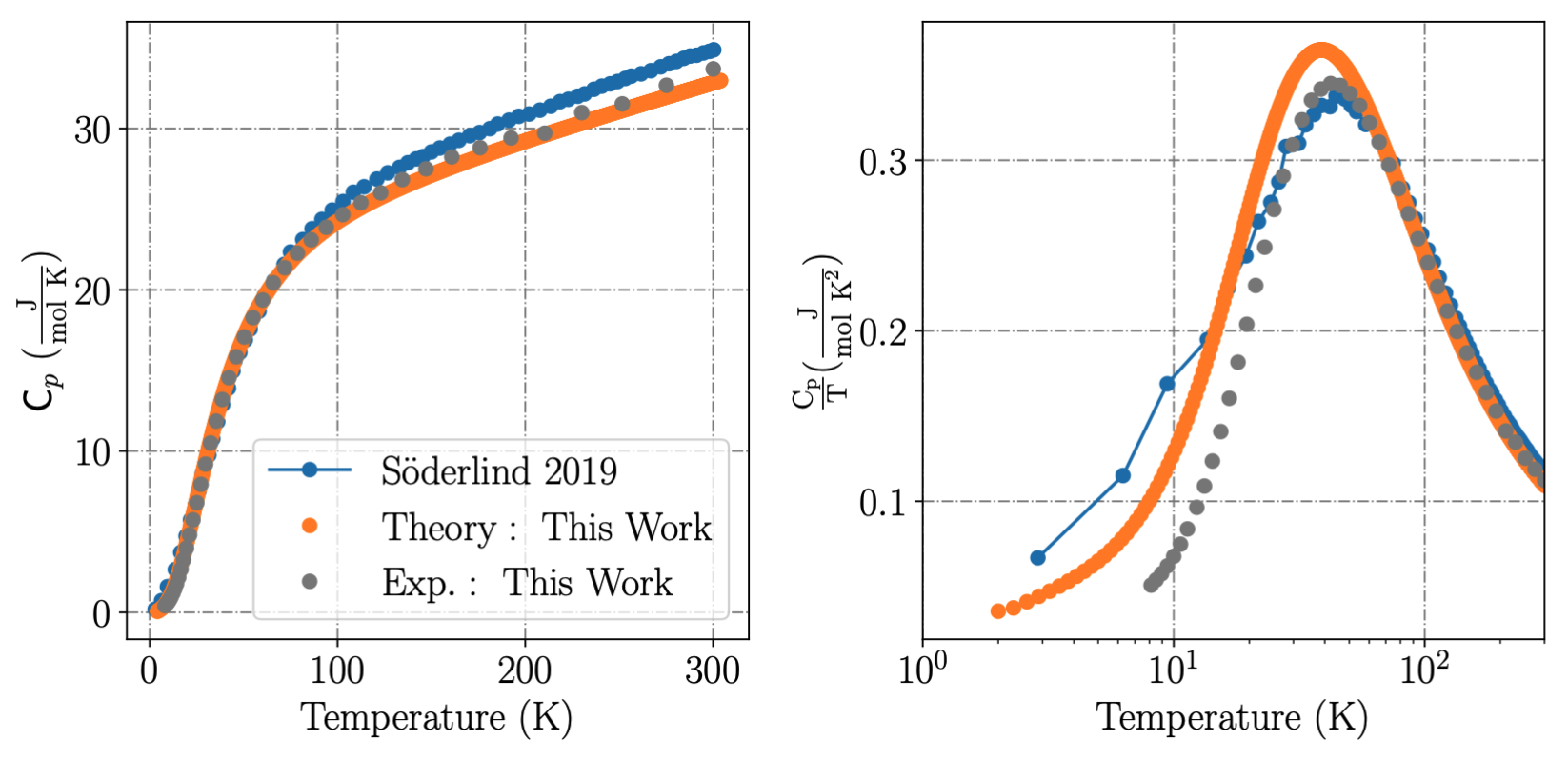}
\caption{Left: Specific heat comparisons between the computed specific heat from the collinear DOS in the literature, the noncollinear DFT of this work, and our experiment.\cite{soderlind_2019} Both theoretical approaches show anomalous specific heat at high temperatures with our DFT approach underestimating the specific heat. Right: The specific heat scaled by temperature reveals low temperature anomalies in the DFT calculations. The low temperature anomaly is explicable from the lack of phonon softening as a function of temperature in our work.}
\label{fig:heat-cap}
\end{figure*}

The electronic atom-projected DOSs, shown in Figure~\ref{fig:e-dos}, show a similar site-dependence to the phonon DOSs.
Each site shows variability in the DOS, indicating site-dependent bonding behavior. 
Notably, Atom 1 has wider peaks in the electronic DOS, specifically above the Fermi energy.
As the site with the greatest number of neighbors at short-bond distances, it is reasonable to expect stronger bonding and higher itineracy.
The stiff bonds in Atom 1's phonon DOS indicate stronger bonding, but this is not clearly supported by the occupied states in the electronic DOS.
For Atoms 4 and 8, the features above the Fermi level indicate the presence of strongly localized states in the system, as seen when using DMFT.\cite{zhu_2013}

\subsection{Thermodynamics}

We show the experimental and theoretical specific heat as a function of temperature in Figure~\ref{fig:heat-cap}.
The specific heat at constant volume is computed as,
\begin{equation}
  C_V(T) = \frac{\partial}{\partial T} \Bigg [\int_0^\infty E \rho(E)n(E,T) dE\Bigg],
\end{equation}
where $\rho(E)$ is the phonon density of states and $n(E,T)$ is the Bose-Einstein distribution.
The specific heat at constant pressure is then obtained by including the thermal expansion and electronic contributions,
\begin{equation}
  C_P(T)=C_V(T)+9B\nu \alpha^2 T+\gamma T,
\end{equation}
where $B$ is the bulk modulus, 55 GPa, $\nu$ is the specific volume, 0.0503 $\frac{cm^3}{g}$, $\alpha$, $54\times10^{-6}\frac{1}{\text{K}}$, is the linear coefficient of thermal expansion, and $\gamma$ is the electronic contribution, 17$\frac{\text{mJ}}{\text{K}^{-2}\text{mol}^{-2}}$.
We use the values from Manley et al. to make the comparison to the experimental and past theoretical specific heats.\cite{manley_2009,soderlind_2019}

With this, we can compare the theoretical specific heats to our experimental measurement.
At high temperatures, the specific heat computed from the noncollinear state underestimates the experimental specific heat, Figure~\ref{fig:heat-cap}.
The collinear state, on the other hand, overestimates the specific heat.
However, neither calculation includes contributions from spin fluctuations.
In the case of the noncollinear state, the underestimate of the specific heat may be explained by this missing contribution.
The collinear state would then overestimate the specific heat.

When the specific heat is scaled by temperature, the low temperature differences between theory and experiment are made apparent.
The primary difference between the theoretical approaches is the tail of the curves at low temperatures. 
The noncollinear specific heat mirrors the distribution of the experimental data as well as the data in the literature studying Schotte-Schotte anomalies in Pu.\cite{harrison_2024}
While we do not take into account phonon softening, the inclusion would shift both curves toward the experiment while maintaining the shape of the distributions.

\begin{figure}
\centering
\includegraphics[width=3in]{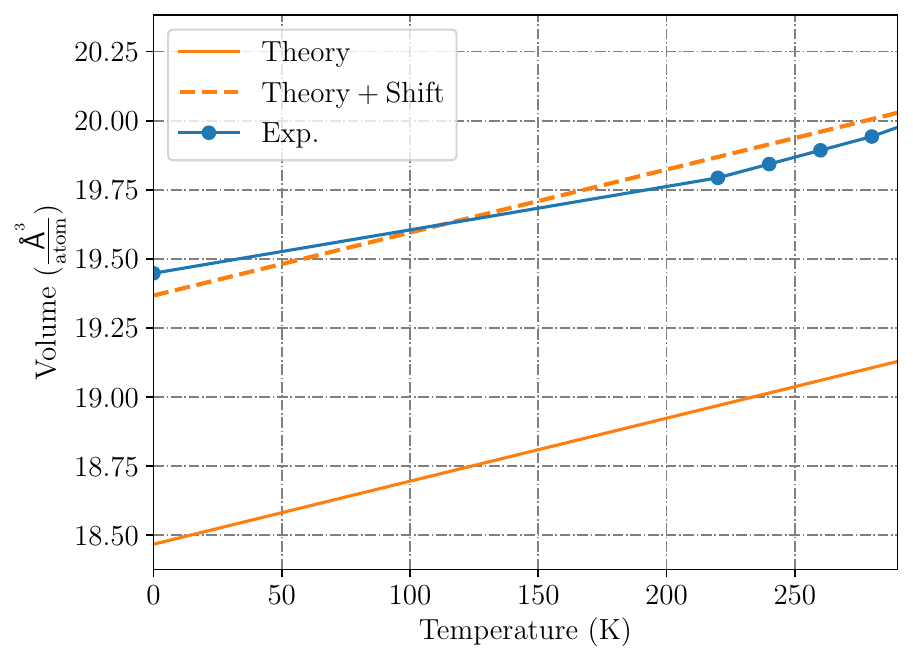}
\caption{Thermal expansion of $\alpha$-Pu calculated in the quasiharmonic approximation. The blue line is the measured atomic volume as a function of temperature. In Pu, PBE underestimates the volume, so the theoretical calculation is shifted, shown in orange, to compare the slope to the experimental slope, shown in green.}
\label{fig:expansion}
\end{figure}

We use the quasiharmonic approximation to compute the thermal expansion of $\alpha$-Pu, Figure~\ref{fig:expansion}.
The orange line shows the computed thermal expansion from PBE.
The curve shows that we systematically underestimate the volume of $\alpha$-Pu, but this behavior is expected from PBE due to its underestimate of electronic correlations in strongly correlated systems.\cite{soderlind_2019}
As such, we shift the PBE result, the dashed orange curve, to compare to the experimental result.\cite{schonfeld_1996}
The inclusion of orbital polarization and the effects of self-irradiation, and defects can alter the slope of the expansion curve, leading to disagreements.
However, we demonstrate that the noncollinear state yields an accurate slope for the thermal expansion when compared to the experiment.

\section{Conclusion}
We explored the state space neighboring the conventional collinear state to identify a low-energy noncollinear state in $\alpha$-Pu.
Using the identified state, we compute the phonon DOS, the atom-projected phonon DOS and the atom-projected electronic DOS.
These results show site-selective interactions that are due to the complex local environments in the system and influence the phononic behavior of the system.
Additionally, we theorize that the noncollinear state space allows for the softening of phonon modes by allowing the state on the atomic site to react to the shortening bonds.

When compared to the the phonon DOS from inelastic x-ray scattering, the noncollinear state's phonon DOS is an improvement over the collinear state.
By performing heat capacity measurements on new material, we make comparisons to our computed heat capacity in an age-controlled setting.
We find that the heat capacity is underestimated at high temperatures, but this is potentially explained by the absence of spin fluctuations, which are of particular importance for the noncollinear state.
As a check on the noncollinear ansatz, we computed the thermal expansion using the quasiharmonic approximation.
While the noncollinear ansatz underestimates the volume at the PBE level, the slope of the curve matches the experimental curve.
In total, these results indicate the utility of the noncollinear ansatz in describing the mechanical properties of $\alpha$-Pu.
However, the effect of the noncollinear states on bonding and the potential application of orbital polarization require further study.

\section{Acknowledgments}
The authors would like to thank P.S. for his advice and encouragement in pursuing this work. 
The authors would like to acknowledge the LDRD project number 20230042DR. 
This research used resources provided by the Los Alamos National Laboratory Institutional Computing Program, which is supported by the U.S. Department of Energy National Nuclear Security Administration under Contract No. 89233218CNA000001.

\bibliographystyle{unsrt}
\bibliography{nonco-alpha}

\end{document}